\documentclass{tlp0}
\usepackage{amssymb}
\usepackage{aopmath}
\usepackage{graphicx}
\usepackage{url}
\usepackage{times}
\usepackage{helvet}
\usepackage{courier}
\usepackage{graphics}
\usepackage{epsf}
\usepackage{rotating}
\usepackage{times}
\usepackage{graphicx}
\usepackage{epsfig}
\usepackage{subfigure}
\usepackage{multirow}
\usepackage{multicol}
\usepackage{comment}

\newtheorem{definition}{Definition}
\newtheorem{example}{Example}

\newcommand{\nnot}{\texttt{not} }
\newcommand{\lparse}{\textsc{lparse} }
\newcommand{\smodels}{\textsc{smodels} }
\newcommand{\smodelsa}{\textsc{smodels$^A$} }
\newcommand{\dlv}{\textsc{dlv} }
\newcommand{\gringo}{\textsc{gringo} }
\newcommand{\clasp}{\textsc{clasp} }
\newcommand{\alparse}{\textsc{alparse} }

\long\def\comment#1{}

\title[Theory and Practice of Logic Programming]
        {Relating Weight Constraint and Aggregate Programs: Semantics and Representation}

  \author[G. Liu and J. You]
         {Guohua Liu and Jia-Huai You\\
         University of Alberta, Edmonton T6G 2R3, Canada\\
         \email{guohua, you@cs.ualberta.ca}}

\begin{document}

\label{firstpage}

\maketitle

\begin{abstract}
Weight constraint and aggregate programs are among the most widely used logic programs with constraints. In this paper, we
relate the semantics of these two classes of programs, namely the stable model semantics for weight constraint programs and the answer set semantics based on
conditional satisfaction for aggregate programs. Both classes of programs are instances of logic programs with constraints, and in particular, the answer set semantics for aggregate programs can be applied to weight constraint programs.
We show that the two
semantics are closely related.
First, we show that for a broad class of weight constraint programs,
called {\em strongly satisfiable programs}, the two semantics
coincide.
When they disagree, a stable model admitted by the stable model semantics may be circularly justified. We show that the gap between the two semantics can be closed by
transforming a weight constraint program to a strongly satisfiable one, so that no circular models may be generated under the current implementation of the stable model semantics. We further demonstrate the close relationship
between the two semantics by
formulating a transformation from weight constraint programs to logic programs with nested expressions which preserves the answer set semantics. Our study on the semantics leads to an investigation of a methodological issue, namely the possibility of compact representation of aggregate programs by weight constraint programs. We show that almost all standard aggregates can be encoded by weight constraints compactly. This makes it possible to compute the answer sets of aggregate programs using
the ASP solvers for weight constraint programs.
This approach is compared experimentally
with the ones where aggregates are handled more explicitly,
which show that the weight constraint encoding of aggregates enables a competitive approach to answer set computation for aggregate programs.

\end{abstract}

\begin{keywords}
Stable model, Weight Constraint, Aggregates, Logic Programs with Constraints.
\end{keywords}

\section{Introduction}
Answer set programming (ASP), namely logic programming under the answer set semantics \cite{g-l-88,niemela-amai}, is a constraint programming paradigm, which has been successfully deployed in many applications \cite{usa-advisor,wu-ieee-tranc-07,caldiran-lpnmr-09,oetsch-lpnmr-09,ielpa-lpnmr-09,delgrande-lpnmr-09,erdem-lpnmr-09}. Recently, ASP was extended to include constraints to facilitate reasoning with sets of atoms. These constraints include weight constraints \cite{simonsAIJ}, aggregates \cite{faber-jelia-04,ferraris-lpnmr-05,pelov-07,son-TPLP-07} and abstract constraints \cite{marek-lpnmr-04,marek-tlp-07,Tru-AAAI-JAIR,son-jair-07,you-lpnmr-07,yidong2009,LPST10}. Among them, weight constraints and aggregates are the most widely used constraints in practice. In this paper, logic programs with weight constraints and aggregates will be referred to as
{\em weight constraint} and {\em aggregate programs}, respectively.

The semantics of weight constraint programs, called the
{\em stable model semantics}, is well established and implemented in a number of ASP solvers \cite{simonsAIJ,bGLM06:JAR,clasp}.
Especially, the results
of the ASP solver competitions \cite{asp-competition,denecker-lpnmr-09} show that \clasp is an efficient solver that implements this semantics.

For aggregate programs, various semantics have been proposed \cite{faber-jelia-04,ferraris-lpnmr-05,pelov-07,son-TPLP-07}. The one proposed in \cite{pelov-07} (previously in \cite{denecker-iclp-01,pelov-lpnmr-04}),
called the {\em ultimate stable semantics}, is based on an iterative construction on partial interpretations.
The same semantics is reformulated by \cite{son-TPLP-07,son-jair-07} and extended to logic programs with arbitrary abstract constraint atoms, which embodies a key concept called {\em conditional satisfaction}.
Since this reformulation is conceptually simpler, as it does not resort to 3-valued logic,
in this paper we call this semantics {\em conditional satisfaction-based}. Among the semantics for aggregate programs,
this semantics is known to be the most
conservative, in the sense that any answer set under this semantics is an answer set under others, but the reverse may not hold. The relationships of these semantics have been studied in \cite{son-jair-07,yidong2009,LPST10}. In this paper, we refer to the semantics based on conditional satisfaction as the {\em answer set semantics}.\footnote{In the literature, {\em stable model} and {\em answer set} are usually interchangeable for logic programs without the ``classic negation''
(see \cite{gelfond}). In this paper, we use them to refer to different semantics.}

Despite the fact that weight constraint and aggregate programs
are among the most popular classes of programs in practice,
the relationship among them has not been fully studied, both in
semantics and in representation.

In this paper, we study the relationship between the stable model semantics and the answer set semantics. We show that for a broad class of weight constraint programs, called {\em strongly satisfiable programs}, the stable model semantics agrees with the answer set semantics. For example, weight constraint programs where weight constraints are upper bound free are all strongly satisfiable. This result is useful in that we are now sure that the known properties of the answer sets also hold for these programs. One important property is that any answer set is a {\em well-supported model} \cite{son-jair-07}, ensuring that any conclusion must be supported by a non-circular justification
in the sense of \cite{fages94}.

Our study further reveals that for weight constraint programs where the stable model and answer set semantics disagree, stable models may be circularly justified.
We then show that the gap between the two can be closed by a transformation, which translates an arbitrary weight constraint program to a strongly satisfiable program so that the answer sets of the original program are exactly the stable models of the translated program.

We further demonstrate the precise difference between the two semantics using a more general logic programming framework, logic programs with nested expressions. We propose yet another transformation from weight constraint programs to logic programs with nested expressions which preserves the answer set semantics. We compare this transformation to the one given in \cite{ferraris-tplp-05}, which is faithful to the stable model semantics. Interestingly, the difference
is small but subtle: given a weight constraint $l [S] u$, where $l$ and $u$ are lower and upper bounds, respectively, and $[S]$ expresses a collection of literals with weights, while in our transformation the satisfaction of the upper bound is interpreted directly as ``less than or equal to $u$'', in \cite{ferraris-tplp-05} the interpretation is by negation-as-failure ``not greater than $u$''.

The observation that the gap between the answer set and the stable model semantics can be closed by a transformation leads to an approach for computing answer sets of aggregate programs using the ASP solvers that implement the stable models semantics of
weight constraint programs. We propose such an approach where aggregate
programs are encoded compactly as weight constraint programs
and their answer sets are computed using a stable model solver. We conducted a series of experiments to evaluate this approach. The results suggest that representing aggregates by weight constraints is a promising alternative to the explicit handling of aggregates in logic programs.

Besides efficiency, another advantage is at the system level: an aggregate language can be built on top of a stable models solver with a simple front end that essentially transforms standard aggregates to weight constraints in linear time. This is in contrast with the state-of-the-art in handling aggregates in ASP, which typically requires an explicit implementation for each aggregate.

The paper is organized as follows. The next section gives preliminary definitions. In Section~\ref{relate} we relate the stable model semantics with the answer set semantics. We first establish a sufficient condition for the two to coincide, and then discuss their differences. In Section \ref{transform}, we present a transformation to close the gap between the two semantics, followed by Section~\ref{aggr-wprogram} where we show how to represent aggregate programs by weight constraint programs.
Further in Section \ref{nested-expression}, to pinpoint the precise difference between the stable model semantics and the answer set semantics for weight constraint programs, by
proposing a transformation from weight constraint programs to logic programs with nested expressions which preserves the answer set semantics, and comparing this with that of \cite{ferraris-tplp-05}.
We implemented a prototype system called \alparse and in
Section~\ref{experiments} we report some experimental results.
Section~\ref{conclusion} concludes the paper.

A preliminary version of this paper has appeared as \cite{liu-iclp-08}. The main extensions here include: (i) Section \ref{nested-expression}, where we propose a transformation from weight constraint programs to logic programs with nested expressions which preserves the answer set semantics \--
this transformation shows exactly what makes the answer set semantics differ
from the stable model semantics; (ii) Section \ref{experiments}, where experiments are expanded
including the benchmarks for aggregate programs
used in the 2007 ASP Solver Competition \cite{asp-competition};
and (iii) the proofs of all the theorems and lemmas.

\section{Preliminaries}
Throughout the paper, we assume a fixed propositional language with a countable set of propositional atoms.

\subsection{Stable Model Semantics for Weight Constraint Programs}
\label{sec-sm-semantics}
 A {\em weight constraint} is of the form
\begin{eqnarray}
l \,[a_1 = w_{a_1},...,a_n = w_{a_n}, \nnot b_1= w_{b_1},...,\nnot b_m= w_{b_m}] \, u
\label{w-form}
\end{eqnarray}
where each $a_i$, $b_j$ is an atom, and each atom and not-atom (negated atom) is associated with a {\em weight}. Atoms and not-atoms are also called {\em literals} (the latter may be emphasized as {\em negative} literals).The literal set of a weight constraint $W$, denoted $lit(W)$, is the set of literals occurring in $W$. The numbers $l$ and $u$ are the {\em lower} and {\em upper bounds}, respectively. The weights and bounds are real numbers. Either of the bounds may be omitted in which case the missing lower bound is taken to be $-\infty$ and the missing upper bound by $\infty$.

A set of atoms $M$ satisfies a weight constraint $W$ of the form (\ref{w-form}), denoted $M\models W$, if (and only if) $l \leq w(W,M) \leq u$, where
\begin{eqnarray}
\label{summation}
\displaystyle w(W,M)=\sum_{a_i\in M}w_{a_i} + \sum_{b_i \not \in M}w_{b_i}
\end{eqnarray}
$M$ satisfies a set of weight constraints $\Pi$ if $M \models W$ for every $W \in \Pi$.

A weight constraint $W$ is {\em monotone} if for any two sets $R$ and $S$, if $R\models W$ and $R \subseteq S$, then $S\models W$; otherwise, $W$ is {\em nonmonotone}. There are some special classes of nonmonotone weight constraints. $W$ is {\em antimonotone} if for any $R$ and $S$, $S\models W$ and $R \subseteq S$ imply $R\models W$; $W$ is {\em convex} if for any $R$ and $S$ such that $R \subseteq S$, if $R\models W$ and $S\models W$ then for any $I$ such that $R \subseteq I \subseteq S$ we have $I \models W$.

A {\em weight constraint program} is a finite set of {\em weight rules} of the form
\begin{eqnarray}
\label{w-rule-form}
W_0 \leftarrow W_1,...,W_n
\end{eqnarray}
where each $W_i$ is a weight constraint.
Given a (weight) rule $r$ of the above form,
we will use $hd(r)$ to denote $W_0$
and $bd(r)$ the conjunction of
the weight constraints in the body of the rule.

We will use $At(P)$ to denote the set of the atoms appearing in a program $P$.

Weight constraint programs are often called {\em lparse programs}, which generally refer to the kind of non-ground, function-free logic programs one can write based on the \lparse syntax.
These programs are grounded before calling an ASP solver.  In this paper, for the theoretical study we always assume a given weight constraint program is ground.

Given a weight constraint program $P$, if
the head of each rule in $P$ is of the form $1~[a=1]~1$ where $a$ is an atom, then $P$ is said to be {\em basic}.
If, in addition, all the weight constraints in the bodies of rules in $P$
are of the form $1~[l=1]~1$, where $l$ is a literal, then we have a
normal program.
We will simply write a weight constraint $1~[l=1]~1$ as $l$, since they are equivalent in terms of satisfaction.

As pointed out by \cite{simonsAIJ}, negative weights and negative literals are closely related in that they can replace each other and that one is inessential when the other is available.
Negative weights can be eliminated by applying the following transformation \cite{simonsAIJ}: Given a weight constraint $W$ of the form (\ref{w-form}), if $w_{a_i} < 0$, then replace $a_i = w_{a_i}$ with $\nnot a_i = |w_{a_i}|$ and increase the lower bound to $l+|w_{a_i}|$ and the upper bound to $u+ |w_{a_i}|$; if $w_{b_i} < 0$, then replace $\nnot b_i = w_{b_i}$ with $b_i = |w_{b_i}|$ and increase the lower bound to $l+|w_{b_i}|$ and the upper bound to $u+ |w_{b_i}|$.

For instance, the weight constraint
\begin{center}
$-1~[a_1 = -1, a_2 = 2, \nnot b_1 = 1, \nnot b_2 = -2] ~1$
\end{center}
can be transformed to
\begin{center}
$2~[\nnot a_1 = 1, a_2 = 2, \nnot b_1 = 1, b_2 = 2] ~4$
\end{center}

Note that this transformation is satisfaction-preserving, in the sense that for any weight constraint $W$ and set of atoms $M$, $M \models W$ iff $M \models W'$, where $W'$ is obtained by applying the transformation.

From now on, we assume that negative weights are always eliminated by the above transformation.

The stable models of weight constraint programs are defined using the reduct of weight constraints, which is defined as follows: The {\em reduct} of a weight constraint $W$ of the form (\ref{w-form}) w.r.t. a set of atoms $M$, denoted by $W^M$, is the weight constraint
\begin{eqnarray}
\label{reduct-weight-constraint}
l'~[a_1=w_{a_1}, ..., a_n=w_{a_n}]
\end{eqnarray}
where $l'=l-\sum_{b_i\not \in M}w_{b_i}$.

Let $P$ be a weight constraint program and $M$ a set of atoms. The reduct of $P$ w.r.t. $M$, denoted $P^M$, is defined by
\begin{eqnarray}
P^M = \{p \leftarrow W_1^M, \ldots, W_n^M ~|~ W_0 \leftarrow W_1, \ldots W_n\in P, ~~~~~~~~~~~~~~~~~~~~~~~~~~~~~\nonumber \\ ~~~~~~~~~~~~~~~~~~~~~~~~~~~~~~~~~~~~~~~~~~~~~~~~~~~~~~~~~~~~~~~~~~~~~~~~p \in lit(W_0) \cap M ~ {\rm and} ~ w(W_i,M) \leq u, ~ {\rm for~all} ~~ i\geq 1\}
\end{eqnarray}

\begin{definition}[{\rm \cite{simonsAIJ}}]
\label{smodels-semantics}
Let $P$ be a weight constraint program and $M \subseteq At(P)$. $M$ is an {\em stable model} of $P$ iff the following two conditions hold:
\begin{enumerate}
\item $M \models P$,
\item $M$ is the deductive closure of $P^M$.
\end{enumerate}
\end{definition}

Note that $P^M$ is a monotone basic weight constraint program. The deductive closure of such a program can be constructed using the operator $T_P$ defined in \cite{Tru-AAAI-JAIR}. Let $P$ be a monotone basic weight constraint program. The operator $T_P$ is defined as
\begin{eqnarray}
T_P(S) = \{h~|~ \exists r \in P  \mbox{ of the form } h\leftarrow bd(r) \mbox{
 and } S \models bd(r)\}
\end{eqnarray}

We note that, for a monotone program $P$, the operator $T_P$ is monotone with respect to $S$. Then we have the lemma below.

\begin{lemma}
\label{sm-fixpoint}
Given a weight constraint program $P$, a set of atoms $M$ is a stable model of $P$ iff $M \models P$ and $M=T_{P^M}^\infty(\emptyset)$.
\end{lemma}

\begin{proof}
Let $P$ be a weight constraint program and $M$ a set of atoms. $P^M$ is a monotone program. The deductive closure of $P^M$ is the least fixpoint of $T_{P^M}(\emptyset)$. Then the lemma follows from Definition \ref{smodels-semantics}.
\end{proof}


\subsection{Answer Set Semantics for Aggregate Programs}

Following \cite{son-TPLP-07}, we define the syntax and semantics for aggregate programs below.

An aggregate is a constraint on a set of atoms taking the form
\begin{eqnarray}
aggr(\{X~|~p(X)\}) \ {\tt op} \ Result
\label{aggregate-form}
\end{eqnarray}
where $aggr$ is an {\em aggregate function}. The standard aggregate functions are those in $\{$SUM, COUNT, AVG, MAX, MIN$\}$. The set $\{X~|~p(X)\}$ is called an {\em intensional set}, where $p$ is a predicate, and $X$ is a variable which takes value from a set $D(X)=\{a_1,...,a_n\}$, called the {\em variable domain}. The relational operator $op$ is from $\{=,\neq,<,>,\leq,\geq\}$ and $Result$ is either a variable or a numeric constant.

The {\em domain} of an aggregate $A$, denoted $Dom(A)$, is the set of atoms $\{p(a)~|~a\in D(X)\}$.  The size of an aggregate is $|Dom(A)|$.

For an aggregate $A$, the intensional set $\{X~|~p(X)\}$, the variable domain $D(X)$, and the domain of an aggregate $Dom(A)$ can also be a multiset which may contain duplicate members.

Let $M$ be a set or multiset of atoms. $M$ is a {\em model} of (satisfies) an aggregate $A$, denoted $M\models A$, if $aggr(\{a~|~p(a) \in M\cap Dom(A)\})~{\tt op}~Result$ holds, otherwise $M$ is not a model of (does not satisfy) $A$, denoted $M \not \models A$.

For instance, consider the aggregate $A=SUM(\{X|p(X)\})\ge 2$, where $D(X)=\{-1,1,1,2\}$. For the sets $M_1=\{p(2)\}$ and $M_2=\{p(-1),p(1)\}$, we have $M_1\models A$ and $M_2 \not \models A$. For the multiset $M_3=\{p(1),p(1)\}$, we have $M_3\models A$.

An aggregate program is a set of rules of the form
\begin{eqnarray}
\label{aggr-rule-form}
h \leftarrow A_1, ..., A_n
\end{eqnarray}
where $h$ is an atom and $A_1, ..., A_n$ are
aggregates.\footnote{In general, the $A_i$'s could also be atoms or negative
atoms. Here we focus on aggregates. The results can be extended to the
general case, where the atoms and negative atoms are treated
exactly the same as that in normal logic programs \cite{son-TPLP-07}.} For a rule $r$ of the form (\ref{aggr-rule-form}), we use $hd(r)$ and $bd(r)$ to
denote $h$ and
the set $\{A_1, ..., A_n\}$, respectively.

The definition of {\em answer set} of aggregate programs is based on the notion of {\em conditional satisfaction}.

\begin{definition}
\label{aggr-cond-sat}
Let $A$ be an aggregate and $R$ and $S$ two sets of atoms. $R$ conditionally satisfies $A$, w.r.t. $S$, denoted $R\models_S A$, iff $R\models A$ and for every set $I$ such that $R\cap Dom(A) \subseteq I \subseteq S \cap Dom(A)$, $I\models A$.
\end{definition}
$R$ conditionally satisfies a set of aggregates $\Pi$ w.r.t. $S$, if $R \models_S A$ for every $A \in \Pi$.

Given two sets $R$ and $S$, and an aggregate program $P$, the operator $K_P(R,S)$ is defined as:\begin{center}
$K_P(R,S)=\{ hd(r)~|~\exists r \in P,~R \models_S bd(r) \}$.
\end{center}

$K_P$ is monotone w.r.t. its first argument, given that the second argument is fixed. Following \cite{son-TPLP-07}, given a set of atoms $M$, the least fixpoint of $K_P$ w.r.t $M$ is defined as $K_P^\infty(\emptyset,M)$, where $K_P^0(\emptyset,M) = \emptyset$ and $K_P^{i+1}(\emptyset,M) = K_P(K_P^i(\emptyset,M),M)$, for all $i \geq 0$ .

\begin{definition} [\rm \cite{son-TPLP-07}]
\label{aggr-ans}
Let $P$ be an aggregate program and $M$ a set of atoms. $M$ is an {\em answer set} of $P$ iff $M$ is a model of $P$ and $M=K_P^\infty(\emptyset, M)$.
\end{definition}

\subsection{Answer Sets of Weight Constraint Programs}

To present our results, it is notationally important to lift the concepts of conditional satisfaction and answer set to weight constraints. Given a weight constraint $W$, the {\em domain of $W$}, denoted $Dom(W)$, is the set $\{a~|~ a\in lit(W) ~~{\rm or}~~ \nnot a \in lit(W)\}$. Let $W$ be a weight constraint and $R$ and $S$ be two sets of atoms. $R$ conditionally satisfies $W$, w.r.t. $S$, denoted $R\models_S W$, if for all $I$ such that $R \cap Dom(W)\subseteq I \subset S\cap Dom(W)$, we have $I\models W$.

First, the answer sets of a basic weight constraint program are defined using the concept of conditional satisfaction.

\begin{definition}
\label{answer-set}
Let $P$ be a basic weight constraint program and $M$ a set of atoms. $M$ is an answer set of $P$ iff $M$ is a model of $P$ and $M=K_P^\infty(\emptyset,M)$.
\end{definition}

Then, following \cite{son-jair-07}, the answer sets of a general weight constraint program are defined as the answer sets of its {\em instances}.

Let $P$ be a weight constraint program, $r$ a rule in $P$ of the form (\ref{w-rule-form}), and $M$ a set of atoms. The {\em instance of $r$ w.r.t. $M$} is

\[ inst(r,M) = \left \{ \begin{array}{ll}
\{a \leftarrow bd(r)~|~ a \in M \cap lit(W_0)\} & \mbox{if $M \models W_0$}\\
\emptyset & \mbox{otherwise}
\end{array}
\right. \]

The {\em instance of $P$ w.r.t. $M$}, denoted $inst(P,M)$, is the program
\begin{eqnarray}
inst(P,M)= \cup_{r\in P} inst(r,M)
\end{eqnarray}

Note that an instance of a program is a basic program.

\begin{definition}
\label{w-program-ans}
Let $P$ be a weight constraint program and $M$ a set of atoms. $M$ is an answer set of $P$ iff $M$ is an answer set of the instance of $P$ w.r.t. $M$.
\end{definition}

In the next section, we will show that, for some weight constraint programs the stable model and the answer set semantics coincide, while, for some others these semantics are different.

Before ending this section, we give a useful proposition, which shows a one-to-one correspondence between the
stable models/answer sets of an arbitrary weight constraint program and those of
its basic program counterparts. This result will be used later in this paper.
\begin{proposition}
\label{basic}
Let $P$ be a weight constraint program and $M$ a model of $P$. Define
\begin{center}
$P' = \{p \leftarrow W_1, \ldots, W_n ~|~ W_0 \leftarrow W_1, \ldots W_n\in P  \mbox{ and } p \in lit(W_0) \cap M\}$,
\end{center}
Then, $M$ is a stable model (resp. answer set) of $P$ iff
$M$ is a stable model  (resp. answer set) of $P'$.
\end{proposition}
\begin{proof}
The correspondence between answer sets follows from Definition \ref{w-program-ans} above. For
the correspondence between stable models,  note that
$P^M = P'^M$.
\end{proof}

\section{Relating Answer Sets with Stable Models}
\label{relate}

In this section, we relate answer sets with stable models. First, we give a sufficient condition under which they agree with each other. Then, we show the difference between these semantics, that is, stable models that are not answer sets may be circular justified, based on a formal notion of circular justification. At the end, we discuss the related justifications in the literature.

\subsection{When Semantics Agree}
\label{coincidence}

We show that for a broad class of weight constraint programs, the stable models are precisely answer sets, and vice versa.

Given a weight constraint $W$ of the form (\ref{w-form}) and a set of atoms $M$, we define $M_a(W) = \{a_i \in M~~|~~a_i \in lit(W)\}$ and $M_b (W) = \{b_i \in M~~|~~\nnot b_i \in lit(W)\}$. Since $W$ is always clear by context, we will simply write $M_a$ and $M_b$.

\begin{definition}
Let $M$ be a set of atoms and $W$ a weight constraint of the form (\ref{w-form}). $W$ is said to be {\em strongly satisfiable by $M$} if $M \models W$ implies that for any $V \subseteq M_b$, $w(W, M \setminus V)\leq u$. $W$ is {\em strongly satisfiable} if for any set of atoms $M$, $W$ is strongly satisfiable by $M$. A weight constraint program $P$ is {\em strongly satisfiable} if every weight constraint that appears in the body of a rule in $P$ is strongly satisfiable.
\end{definition}

Intuitively, a strongly satisfiable weight constraint is a weight constraint whose upper bound is large enough to guarantee that, if a set of atoms satisfies the constraint, then any of its subset also satisfies the constraint.

Strongly satisfiable programs constitute a nontrivial class of programs. In particular, weight constraints $W$ that possess one of the following syntactically checkable conditions are strongly satisfiable.

\begin{itemize}
\item
$lit(W)$ contains only atoms;
\item
$\sum_{i=1}^n w_{a_i} + \sum_{i=1}^m w_{b_i} \leq u$.
\end{itemize}

Note that upper-bound free weight constraints satisfy the second condition above.
\begin{example}
The following constraints are all strongly satisfiable:
$$
\begin{array}{ll}
1~[a=1,b =2]~2
\\
1~[a=1, \nnot b =2]~3
\\
1~[a=1,\nnot b =2]
\end{array}
$$
But the weight constraint
$$1~[a=1,\nnot b =2]~2$$
is not, since it is satisfied by $\{a,b\}$ but not by $\{a\}$. \mathproofbox
\end{example}

Strongly satisfiable weight constraints are not necessarily convex or monotone.
\begin{example}
\label{strong-satisfiable-not-convex}
Let $A$ be the
following weight constraint
$$2[a=1, b=1, \nnot c=1]$$
Since $A$ is upper bound free,
it is
strongly satisfiable. But
$A$ is neither monotone nor convex, as $\{a\} \models A$, $\{a,c\} \not \models A$, and $\{a,b,c\} \models A$. \mathproofbox
\end{example}

We show that the stable model semantics coincides with the answer set
semantics for strongly satisfiable programs. We need a lemma.

\begin{lemma}
\label{key-lemma}
Let $W$ be a weight constraint of the form (\ref{w-form}), and $S$ and $M$ be sets of atoms such that $S \subseteq M$. Then,
\begin{itemize}
\item
[{\rm (i)}] If $S \models_{M} W$ then $S\models W^M$ and $w(W,M) \leq u$.
\item
[{\rm (ii)}] If $S\models W^M$ and $W$ is strongly satisfiable by $M$, then $S \models_{M} W$.\end{itemize}
\end{lemma}

\begin{proof}
{\rm (i)}
We prove it by contraposition. That is,
we show that if $w(W,M) > u$ or
$S \not \models W^M$, then $S \not \models_{M} W$.
The case of $w(W,M) > u$ is simple, which leads to $M \not \models W$ hence
$S \not \models_M W$.

Assume $S \not \models W^M$. By definition, the lower bound is violated, i.e.,
$w(W^M,S) < l'$, where
$l' = l - \sum_{b_i\not \in M}w_{b_i}$.
Let $I = I_a \cup I_b$, where $I_a = S_a$ and $I_b = M_b$.
Since $w(W^M,S) = w(W^I,S)$ and $w(W^M,S) < l'$, we have
$w(W^I,S) < l'$. Then, from $I_b = M_b$ and the assumption
$S \not \models W^M$, we get
$S \not \models W^I$. It then follows from $I_a = S_a$ that
$I \not \models W$.
By construction, we have
$S \cap Dom(W) \subseteq I \subseteq M\cap Dom(W)$, and therefore we conclude
$S \not \models_M W$.

\vspace{.1in}
\noindent
{\rm (ii)}
Assume
$S \not \models_M \! W$ and $W$ is strongly satisfiable by $M$. We show $S \not \models W^M$. We have either $S \models W$ or $S \not \models W$. If
$S \not \models W$ then clearly $S \not \models W^M$. Assume
$S \models W$.
Then from $S \not \models_M W$, we have
$\exists I$, $S \cap Dom(W)\subset I \subseteq M\cap Dom(W)$, such that
$I \not \models W$. Since $W$ is strongly satisfiable by $M$,
if $M \models W$ then
for any $R = M \setminus \!\! V$, where $V \subseteq M_b$,
$w(W,R) \leq u$. Assume $M \models W$.
Let $R$ be such that $R_b = I_b$ and
$I_a \subseteq
R_a$. It is clear that
$w(W,R) \leq u$ leads to $w(W,I) \leq u$.
Thus, since $M \models W$,
that
$I \not \models W$ is
due to the violation of the lower bound, i.e.,
$w(W,I) < l$.

Now consider $I' = S_a \cup M_b$, i.e.,
we restrict $I_a$ to $S_a$ and expand $I_b$
to $M_b$.
Note that by construction, it still holds
that $S \cap Dom(W) \subset I' \subseteq M$.
Clearly, that $I \not \models W$ leads to
$I' \not \models W$, which is
due to the
violation of the lower bound, as $w(W,I') \leq w(W,I)$, i.e., we have
$w(W,I') < l$.
By definition, we have
$w(W^{I'},I') < l'$, where $l' = l -
\sum_{b_i\not \in I'}w_{b_i}$.  Note that since $I'_b = M_b$, we have
$l' = l -
\sum_{b_i\not \in M}w_{b_i}$.
Since $I'_a = S_a$, it follows that
$w(W^{I'},S) < l'$. Now since $W^{I'}$ is precisely the same constraint
as $W^M$,
we have $w(W^{I'},S) =
w(W^M,S)$, and therefore
$w(W^M,S) < l'$. This shows $S \not \models W^M$.
\end{proof}

\begin{theorem}
\label{smodels-son}
Let $P$ be a weight constraint program and $M \subseteq At(P)$.
Suppose for any weight constraint $W$ appearing in the body of a rule in $P$, $W$ is strongly satisfiable by $M$. Then, $M$ is a stable model of $P$ iff $M$ is an answer set of $P$.
\end{theorem}

\begin{proof}
Due to Proposition \ref{basic},
we only need to prove the claim for basic weight constraint programs.

Assume $P$ is a basic weight constraint program and $M$ a model of $P$
such that all weight constraints in $P$ are strongly satisfiable by $M$.
It suffices to prove that for any positive integer $k$, $T_{P^M}^k(\emptyset)=K_P^k(\emptyset, M)$, by induction on $k$.

\vspace{.1in}
\noindent
{\em Base case}: $k=0$. We have $T_{P^M}^k(\emptyset)=K_P^k(\emptyset, M)=\emptyset$.

\vspace{.1in}
\noindent
{\em Induction Step}: Assume, for any $k > 0$,
$T_{P^M}^k(\emptyset)=K_P^k(\emptyset, M)$, and prove that
 $T_{P^M}^{k+1}(\emptyset)=K_P^{k+1}(\emptyset, M)$.
Let $a$ be an atom such that $a \not \in T_{P^M}^{k}(\emptyset)$ and
$a \in T_{P^M}^{k+1}(\emptyset)$.
Then there exists a rule $r \in P$ such that $a \in lit(hd(r))$ and $ T_{P^M}^k(\emptyset) \models W^M$, for each $W \in bd(r)$.
It then follows from part (ii) of Lemma \ref{key-lemma} that
$T_{P^M}^{k}(\emptyset) \models_M W$.
Then $K_P^{k}(\emptyset, M) \models_M W$ by the induction hypothesis. So, $a \in K_P^{k+1}(\emptyset, M)$. Thus $T_{P^M}^{k+1}(\emptyset) \subseteq K_P^{k+1}(\emptyset, M)$. Similarly, we can
show  $K_P^{k+1}(\emptyset, M) \subseteq T_{P^M}^{k+1}(\emptyset)$ using
part (i) of Lemma \ref{key-lemma}.
Thus  $T_{P^M}^{k+1}(\emptyset)=K_P^{k+1}(\emptyset, M)$.

We therefore conclude $T_{P^M}^\infty(\emptyset)=K_P^\infty(\emptyset, M)$.
\end{proof}

The following theorem follows from
Theorem~\ref{smodels-son} and the definition of strongly satisfiable
programs.
\begin{theorem}
\label{ss-program-answer-set}
Let $P$ be a strongly satisfiable weight constraint program and $M \subseteq At(P)$. $M$ is a stable model of $P$ iff $M$ is an answer set of $P$.
\label{coincide}
\end{theorem}

\subsection{When Semantics Disagree}
\label{circularity}

It has been shown that for
logic programs with arbitrary abstract constraints the semantics based on conditional satisfaction are the most conservative in that the answer sets under this semantics are answer sets/stable models of a number of other semantics
\cite{son-jair-07}. It is then expected that the same holds true for
programs with concrete constraints such as weight constraints.

\begin{theorem}
Let $P$ be a weight constraint program.
Every answer set of $P$ is a stable model of $P$, but the converse does not hold.
\end{theorem}

\begin{proof}
Let $M$ be a set of atoms.
Using part (i) of Lemma \ref{key-lemma}, it is easy to show by induction
that for
any positive integer
$k$, we have $K_P^k(\emptyset, M) \subseteq T_{P^M}^k(\emptyset)$.
For the converse, see the counterexample in Example~\ref{key-example} below.
\end{proof}

Question arises as why
some stable models are not answer sets. Later in
Section~\ref{nested-expression}, we will give a technical answer to this
question.  Here,
we suggest that
in these extra stable models
there may exist circular justifications.
Consider the following example.

\begin{example}
\label{key-example}
Let $P$ be a single-rule program:
\begin{eqnarray}
a \leftarrow [\nnot a = 1]~ 0
\end{eqnarray}
Let $M_1=\emptyset$ and $M_2=\{a\}$. The weight constraint $[\nnot a = 1]~ 0$ in $P$ is not strongly satisfiable, since although $M_2$ satisfies the upper bound, its subset $M_1$ does not.
Both $M_1$ and $M_2$ are stable models by Definition \ref{smodels-semantics}.
Note that this is because $P^{M_1} = \emptyset$ and
$P^{M_2} = \{a \leftarrow \}$.
But, $M_1$ is an answer set and $M_2$ is not, by Definition
\ref{w-program-ans}.

The reason that $M_2$ is not an answer set of $P$ is due to the fact that $a$ is derived by its being in $M_2$. This kind of circular justification can be seen more intuitively below using equivalence substitutions.
\begin{itemize}
\item The weight constraint is substituted with an equivalent aggregate:
$$a \leftarrow COUNT(\{X ~|~ X \in D\}) = 1$$
where $D =\{a\}$.
\item The weight constraint is transformed to an equivalent one without negative literal, but with a negative weight,
according to \cite{simonsAIJ}:
$$a \leftarrow [a = -1]{-1}$$
\item The weight constraint is substituted with an equivalent {\em abstract constraint atom} \cite{marek-AAAI-04}\footnote{An {\em abstract constraint atom} is a pair
$(D,C)$, where $D$ is a finite set of ground atoms called the
{\em domain}, and $C$ is a collection of subsets
of $D$ called {\em admissible solutions}. In this example, the set $I= \{a\}$ satisfies the abstract constraint atom, since the admissible solution $\{a\}$ in it is satisfied by $I$.
}:
$$a \leftarrow (\{a\}, \{\{a\}\})$$
\end{itemize}

For the claim of equivalence, note that for any set of atoms $M$, we have:
$M \models [\nnot a = 1]0$ iff $M \models [a = -1]{-1}$ iff
$M \models COUNT(\{X~|~X\in D\}) = 1$ iff
$M \models (\{a\}, \{\{a\}\})$. \mathproofbox
\end{example}

\comment{
The type of circular justification observed here is similar to ``answer sets by reduct'' in dealing with nonmonotone abstract constraint atoms \cite{son-jair-07}.}

 For logic programs with abstract constraint atoms, it is often said that
all of the major semantics coincide for programs with monotone
constraints.
For example, this is the case for the semantics proposed in
\cite{faber-jelia-04,marek-lpnmr-04,ferraris-lpnmr-05,marek-tlp-07,Tru-AAAI-JAIR,son-jair-07,yidong2009,LPST10}.
What is unexpected is that this is not the case for the stable model semantics for weight constraint programs. By the standard definition of monotonicity,
the constraint $[\nnot a = 1]~ 0$ is actually monotone!

One may think that the culprit for $M_2$ above is because it is not a minimal model. However, the following example shows that stable models that are minimal models may still be circularly justified.

\begin{example}
Consider the following weight constraint program $P$ (obtained from the one in Example~\ref{key-example} by adding the second rule):
\begin{eqnarray}
a \leftarrow [\nnot a = 1]~ 0\\
f \leftarrow \nnot f, \nnot a
\end{eqnarray}
Now, $M =\{a\}$ is a minimal model of $P$, and also a stable model of $P$, but clearly $a$ is justified by its being in $M$.
\mathproofbox
\end{example}

We now give a more formal account of {\em circular justification} for stable models, borrowing the idea of {\em unfounded sets} previously used for normal programs \cite{g-r-s-91} and logic programs with monotone and antimonotone aggregates \cite{calimeri-faber-ijcai-05}.
\begin{definition}
\label{cj}
Let $P$ be an weight constraint program and $M$ a stable model of $P$. $M$ is said to be {\em circularly justified}, or simply {\em circular}, if there exists a non-empty set $U\subseteq M$ such that $\forall \phi \in U$, $M\setminus \! U$ does not satisfy the body of any rule $r \in P$ such that
 $\phi \in lit(hd(r))$. Otherwise $M$ is said to be {\em non-circular}.
\end{definition}

\begin{proposition}
\label{as-not-cj}
Let $P$ be a weight constraint program and $M$ a stable model of $P$. If $M$ is an answer set of $P$, then $M$ is not circular.
\end{proposition}
\begin{proof}
Let $P$ be a weight constraint program and $M$ an answer set of $P$.
Assume $M$ is circular.
Then there exists a non-empty subset $U \subseteq M$ such that
$\forall \phi \in U$, $M\setminus U$ does not satisfy the body of any rule $r \in P$ such that
 $\phi \in lit(hd(r))$. By a simple induction on the construction of
$K_P^\infty(\emptyset, M)$, it can be shown that for each of such
$\phi$, we have $\phi \not \in K_P^\infty(\emptyset, M)$. This contradicts
the assumption that $M$ is an answer set.
\end{proof}

Example \ref{key-example} shows that extra stable models (the stable models that are not answer sets) of a program may be circular. However, not all extra stable models are necessarily circular, according to Definition \ref{cj}.
Therefore, the notion of circularity given in
Definition \ref{cj} only serves as a partial characterization of
circular justification.

\begin{example}
\label{non-cj-extra}
Consider a weight constraint program $P$ that consists of the following three rules.
\begin{eqnarray}
b & \leftarrow & 1[\nnot b=1] \label{rule-2} \\
b & \leftarrow & [\nnot b=1]0 \label{rule-3}
\end{eqnarray}
$M=\{b\}$ is a stable model but not an answer set of $P$.
However, it can be verified that $M$ is not circular under
Definition \ref{cj}: $b$ can be derived from the first rule if we don't have $b$, and by the second rule if we do.
\mathproofbox
\end{example}

We shall comment that other forms of non-circular nature of answer sets have been formulated in different ways,
e.g., by the existence of a level mapping \cite{son-jair-07} and by a translation of a constraint to sets of solutions \cite{pelov-asp-03}.

\comment{
From the computational point of view \cite{LPST10}, when we have $a$ and don't have $b$, the body of the rule (\ref{rule-2})
is satisfied which derives $b$. Then the body of the rule (\ref{rule-3})
becomes satisfied, under which $\{a,b\}$ is
stabilized.

On the other hand, the reduct $P^M$ consists of
$$
\begin{array}{ll}
a  \leftarrow  \\
b  \leftarrow  2[a=1]\\
b  \leftarrow  [a=1]
\end{array}
$$
Its deductive closure is $\{a,b\}$ which is thus a stable model.
Since $2[a=1]$ is always false,
it is the 3rd rule above
that made reasoning circular:
$b$ is supported by the stable model due to its being in the model
(so that the reduct of the weight constraint $[a=1, \nnot b=1]1$ becomes
$[a=1]$).

Definition \ref{cj} cannot capture this circular reasoning,
since if $U = \{b\}$, the body of rule (\ref{rule-2}) becomes satisfied.
It appears that circular reasoning is a dynamic phenomenon, which cannot be
captured by syntactic characterizations like unfoundedness in
Definition \ref{cj}.
}

\subsection{Other Justifications}
A semantics is a formal account of intuitions of what justifications for
atoms in a stable model ought to be. For weight constraint programs, there
are different intuitions.

Consider the single-rule program $P$ in Example \ref{key-example} again
\begin{eqnarray}
a \leftarrow [\nnot a=1]0
\end{eqnarray}
One possible interpretation of how $a$ is ``justified'' to be in a stable model is by using the transformation proposed in \cite{marek-tlp-07}. By the transformation, $P$ is translated to $P^*$, which consists the following rules:
\begin{eqnarray}
a & \leftarrow & [b=1]0\\
b & \leftarrow & 0[a=1]0
\end{eqnarray}
It can be verified that the set $M=\{a\}$ is a stable model of $P^*$. The justification of $a$ is: given $M$, $b$ cannot be in any stable model of $P^*$ by the second rule, then $a$ can be derived by the first rule. Note that the above transformation introduces the new atom $b$ in the translated program $P^*$. It assumes that $\nnot a$ implies something new to the original program. The justification of atom $a$ depends on the truth status of the new atom $b$. Whether such a justification is intuitive seems arguable.


Another justification of $a$ is by transforming the program $P$ to a program with nested expressions \cite{ferraris-tplp-05}. We will discuss this in more details later in Section \ref{relation-fl-translation}.

For this example, we also note that the set $M=\{a\}$ is not an answer set under any semantics based on computations studied in \cite{LPST10}.

\section{Transformation to Strongly Satisfiable Programs}
\label{transform}
We show that the gap between the answer set semantics and the stable model semantics can be closed by a transformation, which translates a weight constraint program to a strongly satisfiable program whose stable models are free of circular justifications. In this way, we are able to apply current ASP systems that implement the stable model semantics to compute answer sets for weight constraint programs. In particular, later on we will use strongly satisfiable programs to represent aggregate programs, so that an implementation of aggregate programs can be realized by an implementation of weight constraint programs.

\subsection{Strongly Satisfiable Encoding}
We present an encoding of a weight constraint, where a weight constraint is represented by two strongly satisfiable weight constraints. This encoding captures conditional satisfaction for weight constraints in terms of standard satisfaction. In other words, for weight constraints, the encoding allows conditional satisfaction to be checked by standard satisfaction.

\begin{definition}
Let $W$ be a weight constraint of the form (\ref{w-form}). The {\em strongly satisfiable encoding} of $W$, denoted $(W_l, W_u)$, consists of the following constraints:
$$
\begin{array}{ll}
W_l: l[a_1=w_{a_1},..., a_n=w_{a_n}, \nnot b_1=w_{b_1}, \nnot b_1=w_{b_m}] \\
\\
W_u: -u+ \sum_{i=1}^{n}{w_{a_i}}+ \sum_{i=1}^{m}{w_{b_i}}[\nnot a_1=w_{a_1},..., \nnot a_n=w_{a_n}, b_1=w_{b_1}, ..., b_m=w_{b_m}]
\end{array}
$$
\end{definition}

Intuitively, $W_l$ and $W_u$ are to code the lower and upper bound constraints of $W$, respectively. It is easy to verify that the encoding is satisfaction-preserving, as shown below.

\begin{lemma}
\label{sat-preserving}
Let $W$ be a weight constraint, $(W_l, W_u)$ be its strongly satisfiable encoding, and $M$ be a set of atoms. $M\models W$ iff $M\models W_l$ and $M\models W_u$.
\end{lemma}

\begin{proof}
The satisfaction of $W_l$ is trivial, since $W_l$ is just
the lower bound part of $W$. We show that $W_u$ is the upper bound part of $W$.
Note that the upper bound part of $W$ is
\begin{eqnarray}
\label{c-1}
{\textstyle \sum_{i=1}^n a_i \cdot w_{a_i} + \sum_{i=1}^m b_i\cdot (- w_{b_i}) \leq u-\sum_{i=1}^m w_{b_i} }
\end{eqnarray}
which is equivalent to
\begin{eqnarray}
\label{c-2}
{\textstyle -u+\sum_{i=1}^m w_{b_i}\leq \sum_{i=1}^n a_i \cdot (-w_{a_i})+\sum_{i=1}^m b_i\cdot w_{b_i}}
\end{eqnarray}
By the transformation that eliminates the negative weights (introduced in Section \ref{sec-sm-semantics}), the constraint (\ref{c-2}) is equivalent to the weight constraint $W_u$.
\end{proof}

Using Lemmas \ref{key-lemma} and \ref{sat-preserving}, we establish the following theorem.

\begin{theorem}
\label{cond-sat-reduct}
Let $W$ be a weight constraint, $(W_l, W_u)$ be the strongly satisfiable encoding of $W$, and $S$ and $M$ be two sets of atoms such that $S\subseteq M$. $S\models_M W$ iff $S \models W_l^M$ and $S \models W_u^M$.
\end{theorem}
\begin{proof}

\noindent ($\Rightarrow$) Since $W_l^M$ is the same as $W^M$,  by part (i) of
Lemma \ref{key-lemma}, we have $S\models W_l^M$. In the following, we show $S \models W_u^M$.

Assume $S\models_M W$ and $S\subseteq M$.
Then, by definition, we have
$S\models W$, and $\forall I$ such that $S \cap lit(W) \subseteq I$ and $I \subseteq M\cap lit(W)$, $I\models W$.
Let $I=I_a \cup I_b$ such that $I_b=S_b$ and $I_a=M_a$.
Under this notation, from the assumption $S\models_M W$ and $S\subseteq M$,
we get
$S \cap lit(W) \subseteq I$ and $I\subseteq M\cap lit(W)$. It follows
that $w(W,I) \leq u$, that is, $\sum_{a_i\in I_a}{w_{a_i}} + \sum_{b_i \not \in I_b}{w_{b_i}} \leq u$. This implies
$\sum_{a_i\in M_a}{w_{a_i}} + \sum_{b_i \not \in S_b} \leq u$, from which
the following inequations can be derived

$$
\begin{array}{ll}
\sum_{i=1}^{n}{w_{a_i}} - \sum_{a_i\not \in M_a}{w_{a_i}} + \sum_{i=1}^{m}{w_{b_i}} - \sum_{b_i \in S_b}{w_{b_i}} \leq u\\
\sum_{b_i \in S_b}{w_{b_i}} \geq -u+\sum_{i=1}^{n}{w_{a_i}}+\sum_{i=1}^{m}{w_{b_i}}-\sum_{a_i\not \in M_a}{w_{a_i}}\\
\sum_{b_i \in S}{w_{b_i}} \geq -u+\sum_{i=1}^{n}{w_{a_i}}+\sum_{i=1}^{m}{w_{b_i}}-\sum_{a_i\not \in M}{w_{a_i}}
\end{array}
$$
The last one shows $S \models W_u^M$.

\vspace{.1in}
\noindent ($\Leftarrow$) Assume $S\models W_l^M$ and $S\models W_u^M$. Since neither has an upper bound, both of them are strongly satisfiable. From part (ii) of Lemma \ref{key-lemma}, we have $S\models_M W_l$ and $S\models_M W_u$. It then follows $\forall I$ such that $S\cap lit(W) \subseteq I$ and $I\subseteq M\cap lit(W)$, $I\models W_l$ and $I\models W_u$. Then by Theorem \ref{sat-preserving}, we have $\forall I$ such that $S\cap lit(W) \subseteq I$ and $I\subseteq M\cap lit(W)$, $I\models W$. This shows $S\models_M W$.
\end{proof}

\subsection{Transformation for the Answer Set Semantics}
Using the strongly satisfiable encoding of weight constraints, a weight constraint program can be translated to a strongly satisfiable program, so that the answer sets of the original program are precisely the stable models of the translated program and vice versa..

\begin{definition}
Let $P$ be a weight constraint program. The {\em strongly satisfiable translation} of $P$, denoted $Tr(P)$, is the program obtained by replacing each $W$ in the body of rules in $P$ by the strongly satisfiable encoding of $W$.
\end{definition}

\begin{theorem}
\label{ans-strong-programs}
Let $P$ be a weight constraint program and $M$ a set of atoms. $M$ is an answer set of $P$ iff $M$ is a stable model of $Tr(P)$.
\end{theorem}

\begin{proof}
Due to Proposition \ref{basic},
we only need to prove the claim for basic weight constraint programs.

Using Theorem \ref{cond-sat-reduct}, we have a one-to-one correspondence between the derivations based on conditional satisfaction (Definition~\ref{answer-set})
and the derivations in the construction of the least model (Definition~\ref{sm-fixpoint}), which can be shown by an easy induction on the length of these constructions.
\end{proof}

\begin{example}
Consider a program $P$ with a single rule:
$$a \leftarrow 0 [\nnot a = 3] 2$$
$Tr(P)$ consists of $$a \leftarrow 0 [\nnot a = 3],~1 [a = 3]$$ The weight constraints in $Tr(P)$ are all upper bound-free, hence $Tr(P)$ is strongly satisfiable. Both $\emptyset$ and $\{a\}$ are stable models of $P$, but $\emptyset$ is the only stable model of $Tr(P)$, which is also the only answer set of $P$.
\mathproofbox
\end{example}

\section{Representing Aggregate Programs by Weight Constraint Programs}
\label{aggr-wprogram}

In this section, we propose an approach to computing the answer sets of an aggregate program. For this, we translate an aggregate program to a strongly satisfiable weight constraint program and then compute
its stable models as answer sets.

\subsection{Aggregates as Weight Constraints}
This section shows that the aggregates can be encoded as weight constraints.
In the following, given sets $M$ and $S$, it is convenient to express the set $M$, restricted to
$S$, as $M_{|S}$ which is defined by $M \cap S$.

\begin{definition}
\label{w-encoding}
Let $A$ be an aggregate in the form (\ref{aggregate-form}). A set of weight constraints $\{W_1, ..., W_n\}$ is an {\em weight constraint encoding} (or {\em encoding}) of $A$, denoted $e(A)$, if for any model $M$ of $A$, there is a model $M'$ of $e(A)$ such that $M'_{|Dom(A)}=M$, and for any model $M'$ of $e(A)$, $M'_{|Dom(A)}$ is a model of $A$.
\end{definition}

We show the encoding of aggregates of the form (\ref{aggregate-form}), where the operator $\tt op$ is $\geq$. The encoding can be easily extended to other relational operators except for the operator $\neq$ (more on $\neq$ later in this section). For example, aggregate $SUM(\{X~|~p(X)\}) > k$ can be expressed as $SUM(\{Y|~p(Y)\}) \geq k+1$.

The encoding works for the aggregates whose variable domain contains only integers. For the aggregates whose variable domain contains real numbers, each real number can be converted to an integer by multiplying a factor. In this case, the $Result$ (in the formula (\ref{aggregate-form})) also needs to be processed accordingly.

For convenience, below we may write negative weights in weight constraints. Recall that negative weights can be eliminated by a simple transformation.

\vspace{.18in}
\noindent{\bf $SUM, COUNT, AVG$} \\
These aggregates can be encoded by weight constraints rather directly.

For instance, aggregate $SUM(\{X~|~p(X)\}) \ge k$ can be represented by
\begin{eqnarray}
\label{sum}
k~[p(a_1)=a_1, ..., p(a_n)=a_n]
\end{eqnarray}
where the domain of the aggregate is $\{p(a_1),...,p(a_n)\}$.

\comment{
Note that, to deal with a multiset in $SUM(\{X~|~p(X)\}) \ge k$, we need to introduce a weight literal for
each value in the variable domain. For example, for the aggregate $A=SUM({X~|~p(X)})>k$ where $X$ is defined
by the multiset $\{p(1),p(1),p(2)\}$,  the aggregate will be encoded by the weight constraint
$k[p_{11}(1)=1, p_{12}(1)=1, p(2)=2]$.
}

Note that, a multiset in $SUM(\{X~|~p(X)\}) \ge k$ can be encoded directly by a weight constraint, since
the latter does not require distinct literals in it.
For example, for the aggregate $A=SUM({X~|~p(X)})>k$ where $X$ is defined by the
multiset $\{p(1),p(1),p(2)\}$,  the aggregate can be encoded by the weight constraint
$k[p(1)=1, p(1)=1, p(2)=2]$.

We note that aggregates $COUNT(\{X~|~p(X)\}) \ge k$ and $AVG(\{X~|~p(X)\}) \ge k$ can be encoded simply by substituting the weights in (\ref{sum}) with $1$ and $a_i-k$ (for AVG the lower bound $k$ is also replaced by zero), respectively. \\

\noindent{\bf $MAX$}\\
Let $A=MAX(\{X~|~p(X)\}) \ge k$ be an aggregate. The idea in the encoding of $A$ is that for a set of numbers $S=\{a_1, ..., a_n\}$, the maximum number in $S$ is greater than or equal to $k$ if and only if

\begin{eqnarray}
\displaystyle \sum_{i=1}^n{(a_i-k+1)}>-\sum_{i=1}^n{|a_i-k+1|}
\label{max-relation}
\end{eqnarray}

For each atom $p(a_i)$, two new literals $p^{+}(a_i)$ and $p^{-}(a_i)$ are introduced. The encoding $e(A)$ consists of the following constraints.
\begin{eqnarray}
\label{max-1}
0~[p(a_i)=-1, p^{+}(a_i)=1, p^{-}(a_i)=1]~0,~1\leq i \leq n\\
\label{max-2}
0~[p(a_i)=-d_i, p^{+}(a_i)=d_i],~1\leq i \leq n\\
\label{max-3}
0~[p(a_i)=d_i, p^{-}(a_i)=-d_i],~1\leq i \leq n\\
\label{max-4}
1~[p(a_1)=d_1, p^{+}(a_1)=d_1, p^{-}(a_1)=-d_1, \nonumber \\...,  p(a_n)=d_n, p^{+}(a_n)=d_n, p^{-}(a_n)=-d_n]\\
\label{max-5}
1~[p(a_1)=1, ..., p(a_n)=1]
\end{eqnarray}
where $d_i=a_i-k+1$.

In the following, for any model $M$ of such an encoding, $a =1$ means $a \in M$ and $a =0$ means $a \not \in M$.

The constraints (\ref{max-1}), (\ref{max-2}) and (\ref{max-3}) are used to encode $|a_i-k+1|$. Clearly, if $a_i>k-1$, we have $p^{+}(a_i)=p(a_i)$ and $p^{-}(a_i)=0$; if $a_i<k-1$, we have $p^{-}(a_i)=p(a_i)$ and $p^{+}(a_i)=0$; and if $a_i=k-1$, we have $p^{+}(a_i)=p(a_i)$ or $p^{-}(a_i)=p(a_i)$.

The constraint (\ref{max-4}) encodes the relation (\ref{max-relation}) and the constraint (\ref{max-5}) guarantees that a model of $e(A)$ is not an empty set.\\

\noindent{\bf $MIN$}\\
Let $A=MIN(\{X~|~p(X)\}) \ge k$ be an aggregate. The idea in the encoding
of $A$ is that for a set of numbers $S=\{a_1, ..., a_n\}$, the minimal number in $S$ is greater than or equal to $k$ if and only if
\begin{eqnarray}
\displaystyle \sum_{i=1}^n{(a_i-k)}=\sum_{i=1}^n{|a_i-k|}.
\label{min-relation}
\end{eqnarray}

Similar to $MAX$, the aggregate $MIN$ can be encoded by the following weight constraints.
\begin{eqnarray}
\label{min-1}
0~[p^{+}(a_i)=1, p^{-}(a_i)=1, p(a_i)=-1]~0,~1\leq i \leq n\\
\label{min-2}
0~[p^{+}(a_i)=d_i, p(a_i)=-d_i],~1\leq i \leq n\\
\label{min-3}
0~[p^{-}(a_i)=-d_i, p(a_i)=d_i],~1\leq i \leq n\\
\label{min-4}
0~[p(a_1)=d_1, p^{+}(a_1)=-d_1, p^{-}(a_1)=d_1, \nonumber \\ ..., p(a_n)=d_n, p^{+}(a_n)=-d_n, p^{-}(a_n)=d_n]~ 0\\
\label{min-5}
1~[p(a_1)=1, ..., p(a_n)=1]
\end{eqnarray}
where $d_i=a_i-k$.

The constraint (\ref{min-1}), (\ref{min-2}) and (\ref{min-3}) are the same to the first three constraints in the encoding of $MAX$ (except for the value of $d_i$), respectively. The constraint (\ref{min-4}) encodes the relation (\ref{min-relation}) and the constraint (\ref{min-5}) guarantees that a model of $e(A)$ is not an empty set.\\

We note that all the encodings above result in weight constraints whose collective size is linear in the size of the domain of the aggregate being encoded.

In the encoding of $MAX$ (similarly for $MIN$), the first three constraints are the ones between the literal $p(a_i)$ and the newly introduced literals $p^{+}(a_i)$ and $p^{-}(a_i)$. We call them {\em auxiliary constraints}. The last two constraints code the relation between $p(a_i)$ and $p(a_j)$, where $i\neq j$. We call them {\em relation constraints}. Let $A$ be an aggregate, we denote the set of auxiliary constraints in $e(A)$ by $a(A)$ and the set of relation constraints by $r(A)$. If $A$ is aggregate $SUM$, $COUNT$, or $AVG$, we have that $r(A)=e(A)$, because no new literals are introduced in their encodings.

\begin{theorem}
\label{encoding-aggr}
The set of weight constraint (\ref{sum}), the set of weight constraints from (\ref{max-1}) to (\ref{max-5}), and the set of weight constraints from (\ref{min-1}) to (\ref{min-5}), are weight constraint encodings (Definition \ref{w-encoding}) of the aggregates $SUM$, $MAX$, and $MIN$, respectively.
\end{theorem}

\begin{proof}
The proof for the encoding of aggregate $SUM$ is straightforward. The proof for the encoding of aggregate $MIN$ is similar to that for $MAX$, which we show below.

Let $M$ be a set of atoms and $M\models A$. Suppose $p(a_1)\in M$ and $a_1 \geq k$. Then, we can construct $M'$ as follows:
\begin{enumerate}[(ii)]
\item $p(a_i) \in M'$ and $p^{+}(a_i) \in M'$, if $p(a_i) \in M$ and $a_i\geq k$;
\item $p(a_i) \in M'$ and $p^{-}(a_i) \in M'$, if $p(a_i) \in M$ and $a_i< k$.
\end{enumerate}

We use $W_1$ to $W_5$ to denote the weight constraints in (\ref{max-1}) to (\ref{max-5}).

It is easy to check that the weight constraints $W_1$, $W_2$, and $W_3$ are satisfied by $M'$. Since $a_1 \geq k$, we have $p(a_1) \in M'$ and $p^{+}(a_1) \in M'$. Therefore $W_4$ and $W_5$ are also satisfied by $M'$. So $M'\models e(A)$.

Let $M'$ be a set of atoms and $M'\models e(A)$. Since $M'$ satisfies $W_1$, $W_2$ and $W_3$, we have $p^{+}(a_i)=p(a_i)$ and $p^{-}(a_i)=0$, for $a_i\geq k$; $p^{-}(a_i)=p(a_i)$ and $p^{+}(a_i)=0$, for $a_i<k-1$; and $p^{+}(a_i)=p(a_i)$ or $p^{-}(a_i)=p(a_i)$, if $a_i=k-1$. Since $M' \models W_4$ and $M' \models W_5$, there must be an $i$, such that $a_i\geq k$ and $p(a_i)=1$. That is, $p(a_i)\in M'_{|Dom(A)}$. Then, we have $M\models A$.
\end{proof}

\subsection{Aggregate Programs as Weight Constraint Programs}
\label{sec-program-translation}
We translate an aggregate program $P$ to a weight constraint program, denoted $\tau (P)$, as follows:

\begin{enumerate}
\item For each rule of the form (\ref{aggr-rule-form}) in $P$, we include
in $\tau(P)$ a weight rule of the form
\begin{eqnarray}
h \leftarrow r(A_1), ..., r(A_n)
\label{transformed-rule-form}
\end{eqnarray}
where $r(A_i)$ is the conjunction of the weight constraints that encode the aggregate $A_i$; and
\item If there are newly introduced literals in the encoding of aggregates, the {\em auxiliary rule} of the form
\begin{eqnarray}
W \leftarrow p(a_i)
\label{auxiliary-rule-form}
\end{eqnarray}
is included in $\tau(P)$, for each auxiliary constraint $W$ of each atom $p(a_i)$ in the aggregates.
\end{enumerate}

Note that a weight constraint program $P$ can be translated to a strongly satisfiable program using the translation $Tr(P)$ given in Section \ref{transform}.

We have the following theorem establishing the correctness of the transformation $\tau$.

\begin{theorem}
\label{thm-translation}
Let $P$ be an aggregate program where the relational operator is not $\neq$. For any stable model $M$ of $Tr(\tau(P))$, $M_{|At(P)}$ is an answer set of $P$.
For any answer set $M$ for $P$, there is a stable model $M'$ of $Tr(\tau(P))$ such that $M'_{|At(P)} = M$.
\end{theorem}

\begin{proof}
The rules of the form (\ref{transformed-rule-form}) are the translated counterpart of the rules in $P$. The auxiliary rules of the form (\ref{auxiliary-rule-form}) are added to enforce the auxiliary constraints.

Note that $Tr(\tau(P))$ is a strongly satisfiable program. Then the theorem follows from Theorem \ref{ans-strong-programs} and Theorem \ref{encoding-aggr}.
\end{proof}

\vspace{.03in}
\noindent
{\bf Remark}
For an aggregate where the relation operator is not '$\neq$', the aggregate can be encoded by a conjunction of weight constraints as we have shown in this section. In this case, logic equivalence leads to equivalence under conditional satisfaction. That is why we only need to ensure that an encoding is satisfaction-preserving.

For an aggregate where the relation operator is '$\neq$', two classes are distinguished. One consists of aggregates of the forms $COUNT(.)\neq k$. For these aggregates, the operator '$\neq$' can be treated as the disjunction of the operators '$>$' and '$<$'. Consider the aggregate $A=COUNT(.)\neq k$. $A$ is logically equivalent to $A_1 \vee A_2$, where $A_1=COUNT(.)>k$ and $A_2=COUNT(.)<k$. Let $R$ and $S$ be two sets of atoms, it is easy to show that $R\models_S A$ iff $R\models_S A_1$ or $R\models_S A_2$. The other class consists of the aggregates of the forms $SUM(.)\neq k$, $AVG(.)\neq k$, $MAX(.)\neq k$, and $MIN(.)\neq k$. For these aggregates, the operator '$\neq$' cannot be treated as the disjunction of '$>$' and '$<$', since the conditional satisfaction may not be preserved. Below is an example.

\begin{example}
Consider the aggregates $A = SUM(\{X~|p(X)\}) \not = -1$, $A_1 = SUM(\{X~| p(X)\}) > -1$, and $A_2 = SUM(\{X~| p(X)\}) < -1$. Note that $A$ is logically equivalent to $A_1 \vee A_2$. Consider $S=\{p(1)\}$ and $M=\{p(1),p(2),p(-3)\}$. While $S$ conditionally satisfies $A$ w.r.t. $M$ (i.e., $S \models_M A$), it is not the case that $S$ conditionally satisfies $A_1$ w.r.t. $M$ or $S$ conditionally satisfies $A_2$ w.r.t. $M$.
\mathproofbox
\end{example}

\section{Transformation to Programs with Nested Expressions}
\label{nested-expression}
In this section, we further relate
answer sets with stable models in terms of
logic programs with nested expressions. We formulate a transformation of weight constraint programs to programs with nested expressions and compare this transformation to the one in \cite{ferraris-tplp-05}. The comparison reveals that the difference of the semantics lies in the different interpretations of the constraint on the upper bounds of weight constraints in a program:
while our transformation interprets it directly, namely as ``less than or equal to'', the one in \cite{ferraris-tplp-05}
interprets it as ``not greater than'', which may create double negations
(the atoms that are preceded by \nnot \nnot) in nested expressions. It is
the semantics of
these double negations
that differentiates
the two semantics.


\subsection{Stable Model Semantics for Programs with Nested Expressions}
In the language of nested expressions, {\em elementary formulas} are atoms\footnote {In the original syntax \cite{lifschitz-nested-1999}, elementary formulas can be atoms or atoms with classical negation $\neg$. The classical negation is irrelevant here.} and symbols $\bot$ (false) and $\top$ (true). {\em Formulas} are built from elementary formulas using the unary connective $\nnot$ and the binary connectives , (conjunction) and ; (disjunction).

A rule with nested expressions is of the form
\begin{eqnarray}
\label{nested-rule-form}
Head \leftarrow Body
\end{eqnarray}
where both $Body$ and $Head$ are formulas.
For a rule $r$ of the form (\ref{nested-rule-form}), we use $hd(r)$ and $bd(r)$ to denote the $Head$ and the $Body$ of $r$, respectively.

A program with nested expressions is a set of rules with nested expressions.

The satisfaction of a formula by a set of atoms $M$ is defined as follows:
\begin{itemize}
\item for a literal $l$, $M \models l$ if $l \in M$
\item $M \models \top$
\item $M \not \models \bot$
\item $M \models (F,G)$ if $M \models F$ and $M \models G$
\item $M \models (F;G)$ if $M \models F$ or $M \models G$
\item $M \models \nnot F$ if $M \not \models F$.
\end{itemize}

The reduct of a formula $F$ with respect to a set of atoms $M$,
denoted $F^M$, is defined recursively as follows:
\begin{itemize}
\item for an elementary formula $F$, $F^M=F$
\item $(F,G)^M=F^M,G^M$
\item $(F;G)^M=F^M;G^M$
\item $(\nnot F)^M = \left \{ \begin{array}{ll}
\bot, & \mbox{if $M \models F$;}\\
\top, & \mbox{otherwise.}
\end{array}
\right.$
\end{itemize}

The reduct of a program $P$ with respect to a set of atoms $M$ is the set of rules
\begin{eqnarray}
Head^M \leftarrow Body^M
\end{eqnarray}
for each rule of the form (\ref{nested-rule-form}) in $P$.

The concept of a stable model is defined as follows.\footnote{In the literature, the term {\em answer set} has been used.
Here,
we use {\em stable model} to avoid possible confusion with the answer sets defined in \cite{son-jair-07}.}
\begin{definition} [\rm \cite{ferraris-tplp-05}]
Let $P$ be a logic program with nested expressions and $M$ a set of atoms. $M$ is a stable model of $P$ if $M$ is a minimal model of $P^M$.
\end{definition}

\subsection{Direct Nested Expression Encoding}
We present a nested expression encoding, called the {\em direct nested expression encoding} of weight constraints. We show that conditional satisfaction of a weight constraint can be captured by the standard satisfaction of the reduct of the encoding of the weight constraint.

In the rest of the paper, we will use the following notation: For a set of literals $X$, we define $X^{+}=\{a~|~a \in X\}$ and $X^{-}=\{a~|~ \nnot a \in X\}$.
\begin{definition}
Given a weight constraint $W$ of the form (\ref{w-form}), the {\em nested expression encoding} of $W$, denoted $NE(W)$, is the formula
\begin{eqnarray}
;_{X\subseteq At(W) {\rm ~and~} X \models W} [(,_{a_i \in X}a_i), (,_{b_i \in (At(W)\setminus X)} \nnot b_i)]
\end{eqnarray}
where
$At(W)$ is the set of atoms in $W$.\footnote{For a more readable notation, let us
use $(\Phi_{cond} Exp)$ for $(;_{cond} Exp)$
and $(\Psi_{cond} Exp)$ for $(,_{cond} Exp)$. Then we can rewrite
this formula as
$$
\Phi_{X\subseteq At(W) {\rm ~and~} X \models W} [(\Psi_{a_i \in X}a_i), (\Psi_{b_i \in (At(W)\setminus X)} \nnot b_i)]$$
}
\end{definition}

Intuitively, $NE(W)$ is a nested expression representing the sets that satisfy $W$.

\begin{example}
Let $W=1[a=1,b=1]1$ be a weight constraint. The subsets of $At(W)$ that satisfy $W$ are $X_1=\{a\}$ and $X_2=\{b\}$. Thus $NE(W)=a, \nnot b; b, \nnot a$.
\mathproofbox
\end{example}

An interesting result of the directed encoding is that
for the resulting nested expression, conditional satisfaction is precisely
the satisfaction of the {\em reduct} of the expression. That is, given
a weight constraint $W$ and two sets of atoms
$S \subseteq M$,  $S \models_M W$ if and only if $S \models NE(W)^M$.

Before giving this result, we prove a lemma first.

\begin{lemma}
\label{nested-lemma-1}
Let $W$ be a weight constraint, $NE(W)$ its nested expression encoding, and $(W_l, W_u)$ its strongly satisfiable encoding. Then, for any two sets of atoms
$S$ and $M$ such that $S\subseteq M$, $S\models NE(W)^M$ iff $S \models W_l^M$ and $S \models W_u^M$.
\end{lemma}

\begin{proof}
In this proof, given a weight constraint $W$ of the form
(\ref{w-form}),
we denote $W^{+} = \{a_1,...,a_n\}$ and $W^{-} = \{b_1,...,b_m\}$;
we express
$NE(W)$ as a disjunction of conjunctions $X_i$'s, i.e.,
$NE(W) = ;_{1\leq i \leq k}X_i$, for some $k \ge 0$.
For notational convenience, such a conjunction may be referred to as
a set as well, i.e.,
given a conjunction $X_i =x_1,...,x_n$, we may use the same $X_i$ to denote
the set $\{x_1,...,x_n\}$, and vice versa. Given a set $S$, we use the notation $\nnot(S) = \{\nnot a~|~a \in S\}$.

Since the proof is mainly about mathematical transformation, for convenience,
we present it as a set of mechanical inferences.

\vspace{.1in}
\noindent ($\Rightarrow$)
We give a detailed proof for $S\models W_l^M$. The proof of $S \models W_u^M$ is similar.

(1) Assume $S \models NE(W)^M$ and $S \subseteq M$.

(2) $\exists X \in \{X_1,...,X_k\}$ such that $S\models X^M$, due to (1) and the definition of $NE(W)$.

(3) $\forall b \in X^{-}$, we have $b \not \in M$, that is, $X^{-} \subseteq {W_l}^{+} \setminus M_b$, due to (2).

(4) $X^{+} \subseteq S$, due to (2).

(5) $w(W_l, X) \geq l$, due to (2).

(6) $\sum_{a_i\in X^{+}}w_{a_i} + \sum_{b \in X^{-}}w_{b_i} \geq l$, due to (5).

(7) $\sum_{a \in X^{+}}w_{a_i} \leq \sum_{a_i\in S} w_{a_i}$, due to (4).

(8) $\sum_{b \in X^{-}}w_{b_i} \leq \sum_{b \not \in M_b}w_{b_i}$ where $M_b=M \cap {W_l}^{+}$, due to (3).

(9) $\sum_{a_i\in S}w_{a_i} + \sum_{b \not \in M}w_{b_i} \geq l$, due to (6), (7) and (8).

(10) $\sum_{a_i\in S}w_{a_i} \geq l-\sum_{b \not \in M}w_{b_i}$, due to (9).

(11) $w(W_l^M, S) \geq l-\sum_{b \not \in M}w_{b_i}$, due to (10).

(12) $S \models W_l^M$, due to (11).

\vspace{.1in}
\noindent ($\Leftarrow$)

(1) Assume $S \models W_l^M$, $S \models W_u^M$ and $S \subseteq M$.

(2) $\sum_{a_i\in S}w_{a_i} + \sum_{b_i\not \in M}w_{b_i} \geq l$, due to (1).

(3) $-\sum_{b_i \in S}w_{b_i} \leq u - \sum_{i=1}^n w_{a_i} - \sum_{i=1}^m w_{b_i} + \sum_{a_i \not \in M} w_{a_i}$, due to (1).

(4) $\sum_{a_i\in M}w_{a_i} + \sum_{b_i \not \in S} w_{b_i} \leq u$, due to (3).

(5) Let $X=S \cup \nnot ((W^{+}\cup W^{-})\setminus M)$.

(6) $\sum_{a_i\in X}w_{a_i} + \sum_{b \not \in X}w_{b_i} \leq \sum_{a_i\in M} w_{a_i}+ \sum_{b_i \not \in S} w_{b_i}$, due to (5) and (1).

(7) $\sum_{a_i\in X}w_{a_i} + \sum_{b \not \in X}w_{b_i} \leq u$, due to (4) and (6).

(8) $l \leq \sum_{a_i\in X}w_{a_i} + \sum_{b \not \in X}w_{b_i} \leq u$, due to (2) and (7).

(9) $X \models W$, due to (8).

(10) $X \in \{X_1,...,X_k\}$, due to (9).

(11) $S \models X^M$, due to (5).

(12) $S \models NE(W)^M$, due to (11).
\end{proof}

\begin{theorem}
\label{nested-theorem-1}
Let $W$ be a weight constraint, and $S$ and $M$ two sets of atoms. Then, $S\models_M W$ iff $S\models NE(W)^M$.
\end{theorem}
\begin{proof}
This follows from Theorem \ref{cond-sat-reduct} and Lemma \ref{nested-lemma-1}.
\end{proof}

\subsection{Transformation for the Answer Set Semantics}
Using the direct nested expression encoding of weight constraints, a weight constraint program can be translated to a program with nested expressions, such that the answer sets of the original program are precisely the stable models of the translated program and vice versa.

\begin{definition}
Let $P$ be a weight constraint program and $r$ a rule of the form (\ref{w-rule-form}) in $P$. The {\em nested expression translation of $r$}, denoted $NE(r)$, is the rule of the form
\begin{eqnarray}
\label{ne-rule-form}
(l_1; \nnot l_1),...,(l_p; \nnot l_p), NE(W_0) \leftarrow NE(W_1),...,NE(W_n)
\end{eqnarray}
where $l_1,...,l_p$ are the positive literals in $W_0$.
\end{definition}

Intuitively, the conjunctive term $(l_1; \nnot l_1),...,(l_p; \nnot l_p)$ represent that we are free to choose the atoms in the head of the rule to include in an answer set.

\begin{definition}
Let $P$ be a weight constraint program. The {\em nested expression translation of $P$}, denoted $NE(P)$, is the program obtained by replacing each rule $r$ in $P$ by $NE(r)$.
\end{definition}

Let $P$ be a weight constraint program, $r$ a rule in $P$, and $M$ a set of atoms. By the definitions of $NE(r)$ and the reduct of a nested expression, we know that the reduct of $NE(r)$ w.r.t. $M$, denoted $NE(r)^M$, is the rule of the form
\begin{eqnarray}
\label{ne_rule_reduct}
l_1,...,l_{p'},NE(W_0)^M\leftarrow NE(W_1)^M,...,NE(W_n)^M
\end{eqnarray}
where $\{l_1,...l_{p'}\}=M \cap lit(W_0)$. We will use this fact in the proofs below.

Our main result is that the answer sets of a weight constraint program coincide with the stable models of its nested expression translation. To establish this, we need to show a one-to-one correspondence between the least fixpoint of the operator $K$ applied on a weight constraint program is the unique minimal model of the reduct of its nested expression translation.

To show the main theorem, we prove two lemmas firstly. One shows the coincidence of the satisfaction of a weight constraint, its nested expression encoding, and the reduct of the nested encoding. Then using this lemma, we show the coincidence of the models of a weight constraint program, its nested expression translation, and the reduct of its nested expression translation. The later lemma helps to establish the correspondence between the least fixpoint of the operator $K$ and the minimal model of the program reduct.

\begin{lemma}
\label{ne-reduct-satisfaction}
Let $W$ be a weight constraint of the form (\ref{w-form}), $NE(W)$ its nested expression encoding, and $M$ a set atoms. Then, $M \models W$ iff $M \models NE(W)$ iff $M \models NE(W)^M$.
\end{lemma}
\begin{proof}
That $M\models W$ iff $M \models NE(W)$ follows directly from the definition of $NE(W)$. We give a proof of the claim $M \models NE(W)$ iff $M \models NE(W)^M$. Suppose $NE(W)=X_1;...;X_k$.

\vspace{.07in}
\noindent
($\Rightarrow$) Since $M\models NE(W)$, there is an $X \in \{X_1,...,X_k\}$ such that $X^{+} \subseteq M$ and $X^{-} \cap M =\emptyset$. By the definition of the reduct of a nested expression, we have $ M \models X^M$ and then $M \models NE(W)^M$.

\vspace{.07in}
\noindent
($\Leftarrow$) Since $M \models NE(W)^M$, there is an $X \in \{X_1,...,X_k\}$ such that $M \models X^M$. By the definition of reduct, we have $X^{+} \subseteq M$ and $X^{-} \cap M =\emptyset$. Therefore $M \models X$ and then $M \models NE(W)$.
\end{proof}

\begin{lemma}
\label{ne-program-reduct-model}
Let $P$ a weight constraint program and $M$ a set of atoms. $M \models P$ iff $M \models NE(P)$ iff $M \models NE(P)^M$.
\end{lemma}

\begin{proof}
The equivalence $M \models P$ iff $M \models NE(P)$ holds simply because
$NE(P)$ is satisfaction-preserving.
We give a proof of the statement $M \models NE(P)$ iff $M \models NE(P)^M$.

\vspace{.07in}
\noindent
($\Rightarrow$) Suppose $M \models NE(P)$. Let $r^M \in NE(P)^M$ be a rule of the form (\ref{ne_rule_reduct}). If $M \models NE(W_i)^M$ for all $1\leq i \leq n$, then by Lemma \ref{ne-reduct-satisfaction}, we have $M \models NE(W_i)$ for all $1\leq i \leq n$. Since $M \models NE(P)$, then $M\models NE(W_0)$. Again by Lemma \ref{ne-reduct-satisfaction}, we know $M \models NE(W_0)^M$. Since $\{l_1,...l_{p'}\}=M \cap lit(W_0)$, we have $M \models hd(r^M)$. As
$r^M$ is arbitrary in $NE(P)^M$, we conclude $M \models NE(P)^M$.

\vspace{.07in}
\noindent
($\Leftarrow$) Suppose $M \models NE(P)^M$. Let $r \in NE(P)$ be a rule of the form (\ref{ne-rule-form}). If $M \models NE(W_i)$ for all $1\leq i\leq n$, then by Lemma \ref{ne-reduct-satisfaction}, we have $M \models NE(W_i)^M$ for all $1\leq i \leq n$. Since $M \models NE(P)^M$, then $M \models NE(W_0)^M$. Again by Lemma \ref{ne-reduct-satisfaction}, we know $M\models NE(W_0)$. It is obvious that $M \models ,_{1\leq i \leq n}(l_i;\nnot l_i)$. So $M\models hd(r)$. As $r$ is any rule in $NE(P)$, we have $M \models NE(P)$.
\end{proof}

\comment{
\begin{lemma}
\label{superset-sat}
Let $F$ be a nested expression and $M$ and $N$ two sets of atoms, such that $M\subseteq N$. If $M \models F^M$, then $N \models F^M$.
\end{lemma}
\begin{proof}
Let  $F$ be a nested expression and $M$ a set of atoms. By the definition of $F^M$, we know that $F^M$ is either $\top$, $\bot$, or a formula consisting of atoms. So, $M\subseteq N$ and $M\models F^M$ imply $N\models F^M$.
\end{proof}
}

\begin{theorem}
\label{nested-theorem}
Let $P$ be a weight constraint program, $NE(P)$ the nested expression translation of $P$, and $M$ a set of atoms. Then, $M$ is an answer set of $P$ iff $M$ is a stable model of $NE(P)$.
\end{theorem}

\begin{proof}
($\Rightarrow$) Let $P$ be a weight constraint program and $M$ an answer set of $P$. Then $M \models P$ and $M=K_{inst(P,M)}^\infty(\emptyset,M)$. By
Lemma \ref{ne-program-reduct-model}, $M \models NE(P)$ and
$M \models NE(P)^M$.
We prove by contradiction that $M$ is a minimal model of $NE(P)^M$.
Suppose for some $M'\subset M$ such that
$M'\models NE(P)$. Note that by Lemma \ref{ne-program-reduct-model}, we also have
$M'\models NE(P)^M$.
Let $a \in M\setminus M'$. Then, there exists a rule $r\in P$ of the form (\ref{w-rule-form}) satisfying that $a\in lit(hd(r))$ and $\exists k$ such that
$K_{inst(P,M)}^k(\emptyset,M)\models_M W_i$, for all $W_i\in bd(r)$. Note that
$K_{inst(P,M)}^k(\emptyset,M) \subseteq M$, hence
Theorem \ref{nested-theorem-1} is applicable, from which we know $K_{inst(P,M)}^k(\emptyset,M) \models NE(W_i)^M$, for all $W_i\in bd(r)$. Since
$M' \models NE(P)^M$, we must have
$a \in M'$, which contradicts to the assumption
that $a \in M \setminus M'$. We therefore conclude
that $M$ is a minimal model of $NE(P)^M$, i.e., $M$ is a stable model of $NE(P)$.

\vspace{.07in}
\noindent
($\Leftarrow$) Let $P$ be a weight constraint program and suppose
$M$ is a minimal model of $NE(P)^M$. By Lemma \ref{ne-program-reduct-model}, $M$ is a model of $P$. By the definitions of the operator $K_P$ and $inst(P,M)$, we have $K_{inst(P,M)}^\infty(\emptyset,M) \subseteq M$.
Let $\Delta = K_{inst(P,M)}^\infty(\emptyset,M)$.
We will prove that $\Delta = M$. For this, let's assume it is not the case, i.e.,
$\Delta \subset M$. Since $M$ is a minimal model of $NE(P)^M$, we have $\Delta \not \models NE(P)^M$. Then there is a rule $r$ of the form
(\ref{w-rule-form})
in $P$ and its corresponding rule $NE(r)$
of the form (\ref{ne_rule_reduct}) in $NE(P)^M$ such that $\Delta \models NE(W_i)^M$
for all $W_i \in bd(NE(r))$, and $\Delta \not \models hd(NE(r))$.
Since $M\models NE(P)^M$, for the rule $NE(r)$, we have $M\models NE(W_0)^M$.
It follows that for some
$L = \{l_1...l_{p'}\} \subseteq M \cap lit(W_0)$,
$L\models NE(W_0)^M$. As
$\Delta \not \models hd(NE(r))$,
it must be the case that $\exists l \in L$ such that $l \not \in \Delta$. By Theorem \ref{nested-theorem-1}, however, that
$\Delta \models NE(W_i)^M$
for all $W_i \in bd(NE(r))$
leads to
$\Delta \models_M W_i$ for all $W_i \in bd(r)$. Then, by the
definitions of the operator $K_P$ and $inst(P,M)$,
we must have
$l \in \Delta$. This is a contradiction. Therefore, it must be the case that
$\Delta = M$, and it follows that $M$ is an answer set of $P$.
\end{proof}

\begin{example}
Consider the program $P$ in Example \ref{non-cj-extra}. Its nested expression translation $NE(P)$ consists of the following rules.
\begin{eqnarray}
a & & \\
b & \leftarrow & a, \nnot b\\
b & \leftarrow & \nnot a, \nnot b; b, \nnot a; a, b
\end{eqnarray}
It can be verified that the only stable model of $NE(P)$ is $\{a\}$, which is also the only answer set of $P$.
\mathproofbox
\end{example}

\subsection{Comparison to Ferraris and Lifschitz's Translation}
\label{relation-fl-translation}
Ferraris and Lifschitz \cite{ferraris-tplp-05} proposed a nested expression encoding of weight constraints. Using this encoding, a weight constraint program can be translated to a program with nested expressions, such that there
is a one-to-one correspondence between the stable models of the weight constraint program and the stable models of its translated program with nested expressions.

The difference between our nested expression translation and
Ferraris and Lifschitz's translation (FL-translation) lies in the interpretation of the upper bound constraint of a weight constraint.

To illustrate this difference, let's denote a weight constraint $W$ of the form (\ref{w-form}) by $l[S]u$, where $[S]=[a_1 = w_{a_1},...,a_n = w_{a_n}, \nnot b_1= w_{b_1},...,\nnot b_m= w_{b_m}]$. We call that $l[S]$ and $[S]u$ the lower bound constraint and upper bound constraint of $W$, respectively. Obviously, the lower bound and upper bound constraints are also weight constraints.

In our translation, the upper bound constraint $[S]u$ is {\em directly} encoded by the sets of atoms that satisfy it. In the
FL-translation, $[S]u$ is encoded as $\nnot u+1[S]$, where $u+1[S]$ is further encoded as the sets of atoms that satisfy the weight constraint $u+1[S]$, possibly creating double negations.

This difference is the {\em only} reason that the stable models of our translated program are the answer sets of the original program while the stable models of the FL-translated program are the stable models of the original program.
It should be clear that the extra
stable models that are not answer sets are created by
double negations generated by the indirect interpretation in the
FL-translation.

We use the following example for an illustration.
\begin{example}
Consider the program $P$ in Example \ref {key-example}, which consist of a single rule
\begin{eqnarray}
a \leftarrow [\nnot a =1]0
\end{eqnarray}
By our translation, $NE(P)$ consists of
\begin{eqnarray}
a \leftarrow a
\end{eqnarray}
The only stable model of $NE(P)$ is $\emptyset$, which is the unique answer set of $P$.
By the FL-translation, the translated program $P'$ is
\begin{eqnarray}
a \leftarrow \nnot~~\nnot a
\end{eqnarray}
The stable models of $P'$ are $\emptyset$ and $\{a\}$. Among them, the set $\{a\}$ is not an answer set, but it is justified by the stable model semantics through the double negation $\nnot~~ \nnot a$.
\mathproofbox
\end{example}

\section{Experiments}
\label{experiments}
The theoretical studies show that an aggregate program can be translated to a weight constraint program whose stable models are precisely the answer sets
of the original program. This leads to a prototype implementation
called \textsc{alparse}\footnote{The name stands for computing aggregate programs by \lparse program solvers.
\lparse is the synonym of weight constraint programs.} to compute the answer sets for aggregate programs. In \textsc{alparse}, an aggregate program is firstly translated to a strongly satisfiable program using the translation given in Section \ref{aggr-wprogram}, then the stable models of the translated strongly satisfiable program are computed using an ASP solver that implements the stable model semantics for weight constraint programs. In the next two subsections, we
use \smodels version 2.34 and \clasp version 2.0.3 respectively
as the underlying ASP solver of \alparse and compare \alparse with the implementations of aggregate programs \smodelsa and \dlv version 2007-10-11.\footnote{We should note that \dlv is a language that allows to express programs belonging to a higher complexity class.}

The experiments are run on Scientific Linux release 5.1 with 3GHz CPU and 1GB RAM. The reported time of \alparse consists of the transformation
time (from aggregate programs to strongly satisfiable programs), the grounding time
(calling to \lparse version 1.1.2 for \smodels and \gringo version 2.0.3 for
\textsc{clasp}), and the search time (by \smodels or \textsc{clasp}).
The time of \smodelsa
consists of grounding time, search time and unfolding time
(computing the solutions to aggregates). The time of \dlv includes
the grounding time and search time (the grounding phase is not separated
from the search in \textsc{dlv}).
All times are in seconds.

\subsection{\alparse based on \smodels}

In this section, we compare our approach with two systems, \smodelsa and \textsc{dlv}.

\vspace{.1in}
\noindent {\bf Comparison with \smodelsa}

We compare the encoding approach proposed in last section to the unfolding approach implemented in the system \smodelsa \cite{smodels-A}.\footnote{The benchmarks and programs can be found at {\tt www.cs.nmsu.edu/$\sim$ielkaban/asp-aggr.html}.} The aggregates used in the first and second set of problems (the company control and employee raise problems) are $SUM$; the third set of problems (the party invitation problems) are $COUNT$, and the fourth and fifth set of problems (the NM1 and NM2, respectively) are $MAX$ and $MIN$, respectively.

The experimental results are reported in Table \ref{smodels-A}, where the ``sample size'' is measured by the argument used to generate the test cases. The times are the average of one hundred randomly generated instances for each sample size. The results show that \smodels is often faster than \textsc{smodels}$^A$, even though both use the same search engine.

Scale-up could be a problem for \textsc{smodels}$^A$, due to exponential blowup. For instance, for an aggregate like $COUNT(\{a  |  a \in S\})\ge k$, \smodelsa would list all {\em aggregate solutions} \cite{son-TPLP-07} in the unfolded program, whose number is $C_{|S|}^k$. For a large domain $S$ and $k$ being around $|S| / 2$, this is a huge number. If one or a few solutions are needed, \alparse takes much less time to compute the corresponding weight constraints than
\textsc{smodels}$^A$.

\vspace{.1in}
\noindent {\bf Comparison with \dlv}

In \cite{armi03} the seating problem was chosen to evaluate the performance
of \textsc{dlv}.\footnote{The program contains disjunctive head, but it can be easily transformed to a non-disjunctive program.} The problem is to generate a sitting arrangement for a number of guests, with $m$ tables and $n$ chairs per table. Guests who like each other should sit at the same table; guests who dislike each other should not sit at the same table. The aggregate used in the problem is $COUNT$.

We use the same setting to the problem instances as in \cite{armi03}. The results are shown in Table \ref{seating}. ``Tables'' and ``Chairs'' are the number of tables and the number of chairs at each table, respectively. The instance size is the number of atom occurrences in a ground program. We report the result of the average over one hundred randomly generated instances for each problem size.

The experiments show that, by encoding logic programs with aggregates as weight constraint programs, \alparse solves the problem efficiently. For large instances, the running time of \alparse is about one order of magnitude lower than that of \dlv and the sizes of the instances are also smaller than those in the language of \textsc{dlv}.

\subsection{\alparse based on \clasp}
\label{experiments-clasp}
We use the benchmarks reported in an ASP solver competition and run all instances for each benchmark.\footnote{We choose the benchmarks that have \dlv programs available. The descriptions of benchmarks and programs can be found at {\tt http://asparagus.cs.uni-potsdam.de/contest/}} In the experiments, we set the cutoff time to 600 seconds. The instances that are solved in the cutoff time are called ``solvable'', otherwise ``unsolvable''. Table \ref{alparse-summary} is a summary of the results. In the table, the ``Time `` is the average running time in seconds for the solvable instances. It can be seen that \alparse constantly outperforms \dlv by several orders of magnitude, except for the benchmark of Towers of Hanoi.

The system \clasp has progressed to support aggregates $SUM$, $MIN$ and $MAX$. The aggregates used in the benchmarks are $SUM$ except for Towers of Hanoi where the aggregate $MAX$ is used. The aggregate $SUM$ is essentially the same as weight constraints. We compare the \clasp programs with the aggregate $MAX$ and the corresponding translated weight constraint programs (note that, the answer sets of this aggregate program correspond to those of the corresponding weight constraint program). The performances of \clasp on these two kinds of programs are similar.

As we have mentioned, the transformation approach indicates that it is important to focus on an efficient implementation of aggregate $SUM$ rather than on the implementation of
other aggregates one by one, since they can be encoded by $SUM$.\footnote{The aggregate $TIMES$ can be translated to $SUM$, using a logarithm transformation, thanks to Tomi Janhunen for the comments during the presentation of \cite{liu-iclp-08}.}

\begin{table}[ht]
\caption{Benchmarks used by \smodelsa}
 \label{smodels-A}
 \begin{minipage}{\textwidth}
     \begin{tabular}{lrrr}
     \hline\hline
        Program & Sample Size & \alparse & \smodelsa\\ \hline
        Company Contr. & 20     & 0.03 & 0.09 \\
        Company Contr. & 40     & 0.18 & 0.36 \\
	Company Contr. & 80     & 0.87 & 2.88 \\
        Company Contr. & 120    & 1.40 & 8.14 \\
	Employee Raise  & 15/5  & 0.01 & 0.69 \\
	Employee Raise  & 21/15 & 0.05 & 4.65 \\
	Employee Raise   & 24/20 & 0.05 & 5.55 \\
	Party Invit.     & 80    & 0.02 & 0.05 \\
	Party Invit.     & 160   & 0.07 & 0.1 \\
	NM1             & 125   & 0.21 & 0.1 \\
	NM1             & 150   & 0.25   & 0.1 \\
	NM2             & 125   & 0.30   & 1.24 \\
	NM2             & 150   & 0.68   & 2.36 \\
      \hline\hline
      \end{tabular}
  \end{minipage}
\end{table}

\begin{table}[ht]
\caption{Seating}
 \label{seating}
 \begin{minipage}{\textwidth}
       \begin{tabular}{lrrrrr}
        \hline\hline
        \multirow{2}{*}{Tables} &
	\multirow{2}{*}{Chairs} &
	\multicolumn{2}{c}{Time} & \multicolumn{2}{c}{Instance Size} \\ 
	  & & \alparse & \dlv & \alparse & \dlv \\ \hline
        3 & 4 & 0.1  & 0.01   & 293 & 248     \\
        4 & 4 & 0.2  & 0.01   & 544 & 490     \\
        5 & 5 & 0.23  & 0.01   & 1213 & 1346   \\
        10 & 5 & 0.30  & 0.27   & 6500 & 7559   \\
        15 & 5 & 0.88 & 1.52  & 18549 & 22049 \\
        20 & 5 & 1.35  & 4.08   & 40080 & 47946 \\
        25 & 5 & 6.19  & 58.29  & 73765 & 88781   \\
        30 & 5 & 10.42 & 110.45 & 12230 & 147567 \\
      \hline\hline
      \end{tabular}
  \end{minipage}
 \end{table}

\begin{table}[ht]
\caption{Benchmarks from ASP Competition}
\label{alparse-summary}
\begin{minipage}{\textwidth}
      \begin{tabular}{lrrrrr}
        \hline\hline
        \multirow{2}{*}{Benchmarks} &  \multirow{2}{*}{Number of Instances} & \multicolumn{2}{c}{Solved Instances}  & \multicolumn{2}{c}{Time}\\
	&   & \alparse & \dlv & \alparse & \dlv \\ \hline	
	15 Puzzle & 11 & 11   & 11  & 0.31  & 1.16   \\
	Schur Number & 5 & 5   & 4  & 0.10 &  0.62 \\
	Blocked N-queens & 37 & 37  & 12 & 8.94 & 328.92 \\
        Wt. Spanning Tree & 30  & 30 & 30  & 0.12 &  0.17\\
	Bd. Spanning Tree & 30 & 30 & 5  &  1.91 & 414.42\\
	Hamiltonian Cycle & 29 & 29 & 29 & 0.84 & 29.22 \\
	Towers of Hanoi & 29 & 29 & 21 & 21.61 & 18.35\\
	Social Golfer   & 168 & 129 &  107 & 1.52 & 14.69 \\
	Wt. Latin Square & 35 & 35 & 18 &  0.03 & 105.01\\
	Wt. Dominating Set & 30 & 23 & 3 & 0.26 & 192.53 \\
	Traveling Sales  & 24 & 24 & 23 &  0.11 & 12.74\\
	Car Sequencing & 54 & 23 & 0 & 0.08 & -- \\
        \hline\hline
      \end{tabular}
      \end{minipage}
\end{table}

\section{Conclusion}
\label{conclusion}
We have shown that for a large class of programs the stable model semantics coincides with the answer set semantics based on conditional satisfaction. In general, answer sets admitted by the latter are all stable models. When a stable model is not an answer set, it may be circularly justified. We have proposed a transformation, by which a weight constraint program can be translated to strong satisfiable program, such that all stable models are answer sets and thus well-supported models.
We have also given
another transformation from
weight constraint programs to logic programs with
nested expressions which preserves the answer set semantics. In conjunction
with
the one given in \cite{ferraris-tplp-05}, their difference reveals precisely
the relation between stable models and answer sets.

As an issue of methodology, we have shown that most standard aggregates can be encoded by weight constraints. Therefore the ASP systems that support weight constraints can be applied to efficiently compute the answer sets of logic programs with aggregates. The experimental results demonstrate the effectiveness of this approach.

Currently, \alparse
does not handle programs with aggregates like
$SUM(.) \neq k$ or $AVG(.) \neq k$, due to the fact that
the complexity of such programs is higher than $NP$. What is the best way to
include this practically requires further investigation.

\bibliographystyle{acmtrans}

\label{lastpage}
\end{document}